\documentclass[%
reprint,
superscriptaddress,
showpacs,preprintnumbers,
amsmath,amssymb,
prb,
]{revtex4-1}
\usepackage{graphicx}
\usepackage{dcolumn}
\usepackage{bm}

\usepackage{graphicx}
\usepackage{dcolumn}
\usepackage{xfrac}
\usepackage{bm}
\usepackage{color}
\usepackage[export]{adjustbox}
\usepackage{hyperref}
\usepackage{tabularx}

\newcommand{\RCL}{$\alpha$-RuCl$_3$}
\newcommand{\CIO}{Cu$_2$IrO$_3$}
\newcommand{\NIO}{Na$_2$IrO$_3$}
\newcommand{\LRO}{Li$_2$RhO$_3$}
\newcommand{\LIO}{$\alpha$-Li$_2$IrO$_3$}

\newcommand{\ALIO}{Ag$_{3}$LiIr$_2$O$_6$}
\newcommand{\ALRO}{Ag$_{3}$LiRh$_2$O$_6$}
\newcommand{\CLIO}{Cu$_{3}$LiIr$_2$O$_6$}

\newcommand{\HLIO}{H$_{3}$LiIr$_2$O$_6$}
\newcommand{\NCTO}{Na$_{2}$Co$_{2}$TeO$_{6}$}
\newcommand{\NZTO}{Na$_{2}$Zn$_{2}$TeO$_{6}$}
\newcommand{\NCSO}{Na$_{3}$Co$_{2}$SbO$_{6}$}

\begin{document}

\title{Introducing the monoclinic polymorph of the Kitaev magnet \NCTO}
\thanks{These authors contributed equally to this work.}

\author{Emilie~Dufault$^\ast$}
\affiliation{Department of Physics, Boston College, Chestnut Hill, MA 02467, USA}

\author{Faranak~Bahrami$^\ast$}
\affiliation{Department of Physics, Boston College, Chestnut Hill, MA 02467, USA}

\author{Alenna~Streeter}
\affiliation{Department of Physics, Boston College, Chestnut Hill, MA 02467, USA}

\author{Xiaohan~Yao}
\affiliation{Department of Physics, Boston College, Chestnut Hill, MA 02467, USA}

\author{Enrique~Gonzalez}
\affiliation{Department of Physics, Boston College, Chestnut Hill, MA 02467, USA}

\author{Qiang~Zhang}
\affiliation{Neutron Scattering Division, Oak Ridge National Laboratory, Oak Ridge, TN 37830, USA}

\author{Fazel~Tafti}
\affiliation{Department of Physics, Boston College, Chestnut Hill, MA 02467, USA}
\email{fazel.tafti@bc.edu}
\begin{abstract}
Recent theoretical studies have suggested that the low-energy Hamiltonian of honeycomb cobaltate systems could be dominated by anisotropic Kitaev interactions. 
Motivated by the theory, a honeycomb layered material \NCTO\ with a hexagonal unit cell has been studied and found to exhibit antiferromagnetic (AFM) ordering at 27~K with two spin reorientation transitions at 15 and 5~K.
Here we report a monoclinic polymorph of \NCTO, also with honeycomb layered structure but with a single AFM transition at 9.6~K and without spin reorientation transitions at lower temperatures. 
Using neutron diffraction, we identify an in-plane zigzag AFM order in the ground-state with the spins canted out of the honeycomb planes and ferromagnetically coupled between them.
The zigzag order is suppressed by a magnetic field of 6~T.
\end{abstract}

\maketitle

\section{\label{sec:INTRO}Introduction}
Establishing a quantum spin-liquid (QSL) phase is highly desired in condensed matter physics, since the non-abelian anyonic excitations of a QSL can be used as qubits for topological quantum computing~\cite{science-qsl-2020,motome_hunting_2020,ARofCMP_Guide_sl-2019,Savary_2017,kitaev_fault-tolerant_2003,nayak_non-abelian_2008}. 
One of the most promising proposals for the QSL phase is the Kitaev model based on anisotropic interactions among spin-$1/2$ particles on a honeycomb lattice~\cite{KITAEV20062}. 
Experimental efforts to materialize the Kitaev model have been largely focused on honeycomb layered structures with heavy transition metals such as \LIO, \NIO, \LRO, \RCL, \CIO, \ALIO, \ALRO, \CLIO, and \HLIO~\cite{PRL_Singh_2012, PRB_kemp_2014, Mehlawat_HC_2017,Mazin_LRO_2013,Todorova_LRO_2011,abramchuk_cu2iro3_2017,Bahrami_PRL_ALIO,Bahrami_PRB_ALIO,Bahrami_ALRO_2022,roudebush2016iridium,kitagawa2018spin,PRB-Eric-CIO-2019,PRB-NMR-ALIO-2021,PRB-ALIO-Hybridization-2021,PRB-pressure-CIO-2021,molecules-FB-2022,de_la_torre_momentum-independent_2023}. 
The choice of $4d$ and $5d$ transition metals (Ru, Rh, Ir) is due to their strong spin-orbit coupling (SOC) that induces anisotropic interactions among pseudospin-$1/2$ ($J_{\text{eff}} = 1/2$) spin-orbital states~\cite{chaloupka_kitaev-heisenberg_2010,takagi2019concept,annurev-conmatphys}. 
Such $J_{\text{eff}} = 1/2$ Kramers doublets originate from the low-spin configuration $t^{5}_{2\textrm g}\,e^0_{\textrm g}$ of the $(4,5)d^{5}$ orbitals of Ru$^{3+}$, Rh$^{4+}$, and Ir$^{4+}$ subjected to octahedral crystal electric field (CEF)~\cite{rau_generic_2014}. 

Recent theoretical studies have suggested that both the anisotropic exchange interactions and Kramers doublets can also be realized in the high-spin configuration $t^{5}_{2g}e^{2}_{g}$ of the $3d^{7}$ orbitals of Co$^{2+}$ and Ni$^{3+}$~\cite{Kitav-heisenberg-3d-2018,theory-Co-Kitaev-2018,stavropoulos_microscopic_2019,KSL-3d-2020}.
The tantalizing possibility of synthesizing Kitaev QSL candidate materials with earth-abundant elements (Co and Ni) instead of precious metals (Ru, Rh, and Ir) led to a surge of activity on such materials as \NCSO\ and \NCTO~\cite{NCSO-zigzag-2016,VICIU-NaCTO-NCSO,yan_magnetic_2019,PRB-CIF-NaCTO-2017,NaCTO_ACS_2019,NCTO_Cm,nature-NCTO-2021,PhysRevMaterials-NaCTO_2022}.
In these compounds, anisotropic interactions stem from a sizable Hund's coupling in the $e_\textrm{g}$ manifold and enhanced SOC effect of the ligands due to proximity of oxygen to heavier Sb or Te atoms~\cite{stavropoulos_microscopic_2019}.

\begin{figure}
\includegraphics[width=0.46\textwidth]{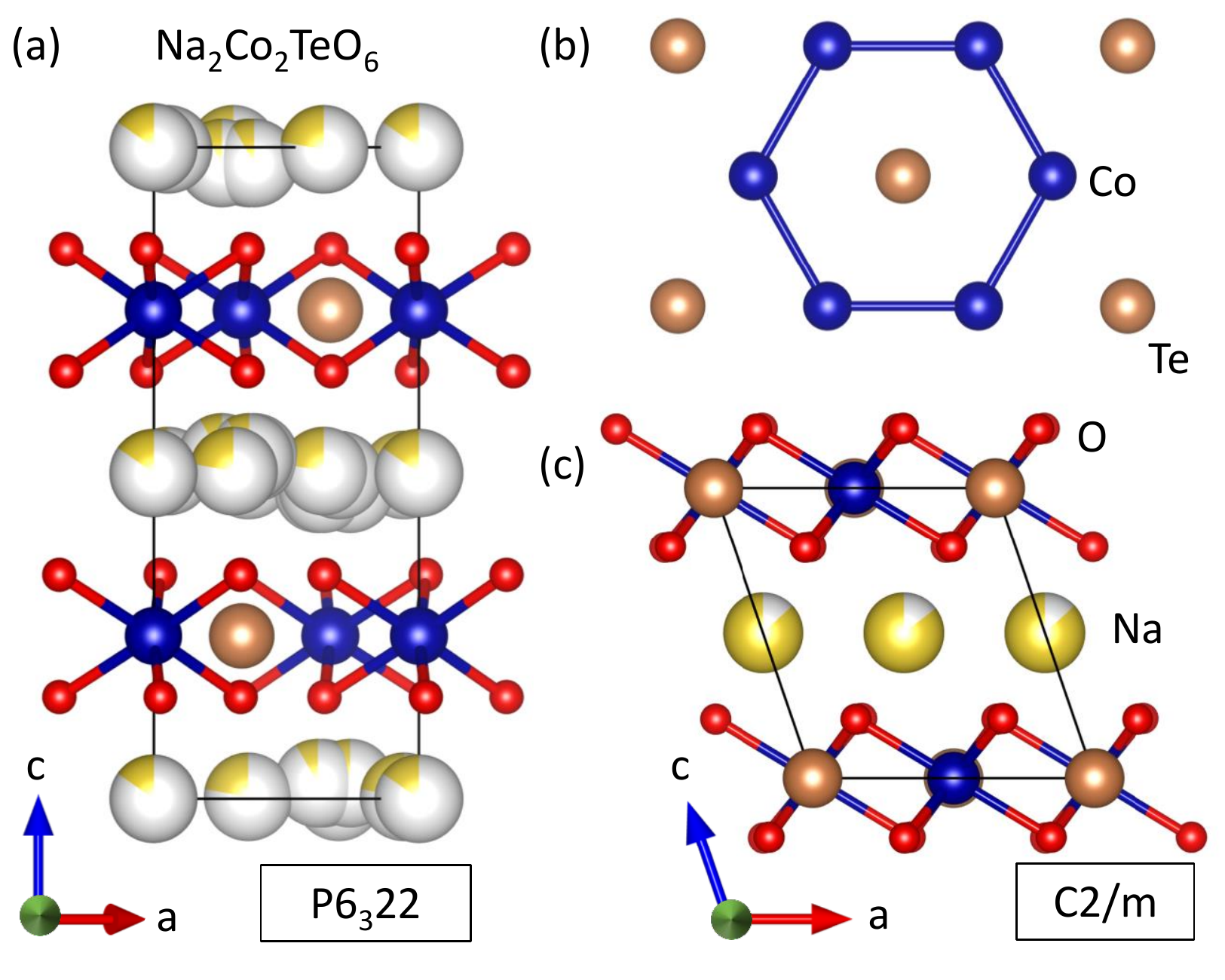}
\caption{\label{fig:CIF} 
(a) The hexagonal polymorph of \NCTO\ has considerable sodium deficiency and site disorder between the layers.
The yellow and white colors show Na occupancy and vacancy, respectively.
(b) Both hexagonal ($P6_322$) and monoclinic ($C2/m$) space groups have honeycomb layers. 
(c) The monoclinic polymorph has less inter-layer sodium disorder.
}
\end{figure}

\NCSO~crystallizes in the monoclinic space group $C2/m$ similar to the iridates.
It shows anti-ferromagnetic (AFM) ordering at $T_\text{N}=8.3$~K with a positive Curie-Weiss temperature $\Theta_{\text{CW}}=+2.2$~K. 
The positive $\Theta_{\text{CW}}$ and a weak hysteresis in $M(H)$ at 2~K despite AFM ordering suggest a competition between ferromagnetic (FM) and AFM interactions in this material~\cite{NCSO-zigzag-2016,VICIU-NaCTO-NCSO,yan_magnetic_2019}. 
\NCTO\ instead crystallizes in the hexagonal space group $P6_{3}22$. 
It undergoes an AFM transition at 27~K followed by two spin reorientation transitions at 15 and 5~K. 
The negative $\Theta_{\text{CW}}=-8.3$~K in polycrystalline samples confirms dominant AFM interactions, unlike competing FM and AFM interactions found in \NCSO~\cite{NaCTO_ACS_2019,NCTO_Cm,nature-NCTO-2021,PhysRevMaterials-NaCTO_2022}. 

Both the monoclinic ($C2/m$) unit cell of \NCSO\ and hexagonal ($P6_{3}22$) unit cell of \NCTO\ posses sodium disorder between the honeycomb layers.
However, there is more disorder in the hexagonal structure because it allows for three inter-layer Wyckoff sites unlike the monoclinic structure with two inter-layer Wyckoff sites.
Such disorder in the inter-layer site occupancy randomizes the position of oxygen atoms and leads to higher levels of bond randomness within the honeycomb layers and stacking faults between them (Fig.~\ref{fig:CIF}a).

In this article, we introduce a monoclinic polymorph of \NCTO\ in the space group $C2/m$, which is structurally similar to \NCSO.
As shown in Fig.~\ref{fig:CIF}, the two-layer monoclinic polymorph reported here has a smaller amount of inter-layer sodium disorder than the three-layer hexagonal polymorph~\cite{PRB-CIF-NaCTO-2017,NaCTO_ACS_2019,nature-NCTO-2021,PhysRevMaterials-NaCTO_2022}. 
Unlike the hexagonal \NCTO\ that has three transitions at 27, 15, and 5~K, the monoclinic polymorph has a single AFM transition at 9.6~K. 
Also, the large splitting between zero-field-cooled (ZFC) and field-cooled (FC) susceptibility in the hexagonal \NCTO, indicative of spin-glass behavior, is absent in the monoclinic polymorph consistent with lower disorder levels.

\section{\label{sec:EXP}Experimental Methods}
Polycrystalline samples of both hexagonal and monoclinic \NCTO\ were synthesized via a solid-state reaction. 
The precursor materials sodium carbonate (99.5$\%$), cobalt oxide (99.7$\%$), and tellurium oxide (99.99$\%$) were mixed and reacted according to the following equation.
\begin{equation}
\label{eq:SYNTHESIS}
3\textrm{Na}_{2}\textrm{CO}_{3} + 2 \textrm{Co}_{3}\textrm{O}_{4} + 3 \textrm{TeO}_{2} \rightarrow 3 \textrm{Na}_{2}\textrm{Co}_{2}\textrm{TeO}_{6}
\end{equation}
The mixture was pressed into a 350 mg pellet, wrapped in a gold foil, and sintered in a capped alumina crucible at 850$^{\circ}$C for 24~h.
It was then cooled to 550$^{\circ}$C and quenched in a dry box.
The hexagonal polymorph was obtained by following Eq.~\ref{eq:SYNTHESIS} strictly, and the monoclinic polymorph was obtained by adding 30\% molar excess of Na$_2$CO$_3$.
Both polymorphs were fairly stable in air and had distinguishable colors of purple (monoclinic) and maroon (hexagonal) as shown in Fig.~\ref{fig:XRD}. 
We also synthesized the non-magnetic analog \NZTO~with a similar approach (using 50\% additional Na$_2$CO$_3$) to subtract the phonon background from the heat capacity data.

Powder x-ray diffraction (PXRD) measurements were performed using a Bruker D8 ECO instrument with a Cu-K$\alpha$ source. 
The FullProf suite~\cite{FullProf} and VESTA software~\cite{Vesta} were used for the Rietveld refinements and crystal visualizations. 
Magnetization and heat capacity measurements were performed using a Quantum Design MPMS3 and PPMS Dynacool, respectively. 
Neutron powder diffraction (NPD) was performed on the time-of-flight (TOF) powder diffractometer POWGEN at the Spallation Neutron Source at Oak Ridge National Laboratory by loading 2.5~g of dried powder into a vanadium sample can and cooling it in an orange cryostat. 
For optimal nuclear and magnetic refinements, two neutron banks with center wavelengths of 1.500~\AA\ and 2.556~\AA\ were selected, respectively, at 100~K and 1.6~K. 
The Fullprof $\bm{k}$-Search software was used to identify the magnetic propagation vector~\cite{FullProf}. 
The Bilbao Crystallographic Server~\cite{perez-mato_symmetry-based_2015} was used for the magnetic symmetry analysis, and GSAS-II~\cite{toby_gsas_2013} was used for the refinements. 

\section{\label{sec:RES}Results and Discussion}

\subsection{\label{subsec:STRUC}Structural Analysis}
\begin{figure}
\includegraphics[width=0.46\textwidth]{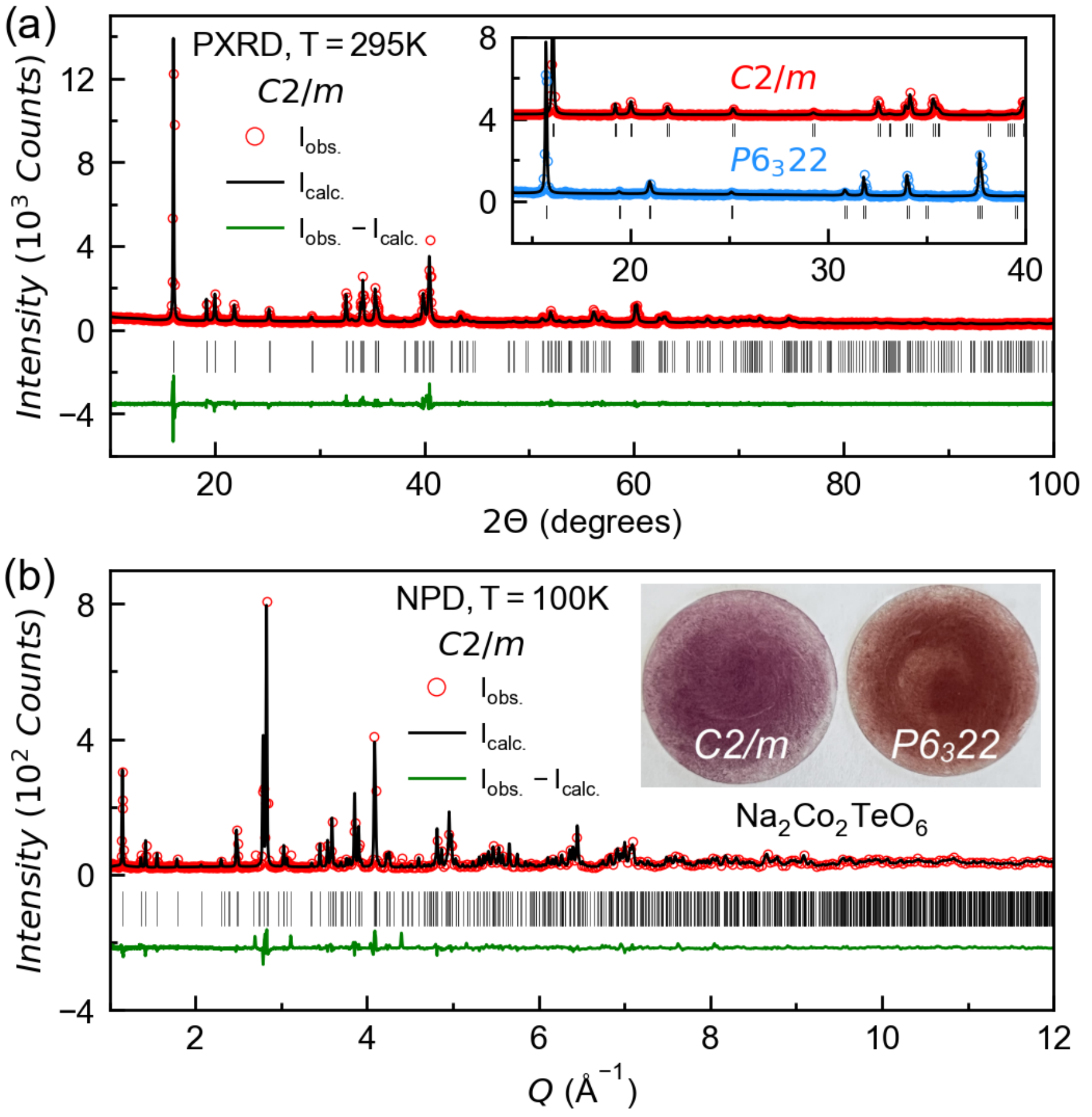}
\caption{\label{fig:XRD} 
(a) Rietveld refinement of the PXRD pattern of monoclinic \NCTO. The inset compares PXRD patterns of the monoclinic ($C2/m$) and hexagonal ($P6_322$) polymorphs. 
(b) Rietveld refinement of the NPD pattern at $T\gg T_{\text{N}}$. The inset compares the colors of the monoclinic (purple) and hexagonal (maroon) polymorphs.
}
\end{figure}
Figures~\ref{fig:XRD}a,b show the PXRD and NPD patterns of the monoclinic polymorph of \NCTO~(red empty circles) with Rietveld refinements in the $C2/m$ space group (black solid lines).
The crystallographic solution confirmed by both PXRD and NPD is visualized in Figs.~\ref{fig:CIF}b,c, and the refinement details are summarized in Appendix~\ref{app:RF}. 
The inset of Fig.~\ref{fig:XRD}a shows visible differences between the PXRD patterns of the monoclinic ($C2/m$) and hexagonal ($P6_322$) polymorphs.
The first peak for the hexagonal compound is located at a lower angle compared to that of the monoclinic compound suggesting a stronger inter-layer connection and smaller inter-layer spacing in the monoclinic polymorph. 
The inset of Fig.~\ref{fig:XRD}b shows that the two polymorphs have different colors. 
As shown in Fig.~\ref{fig:CIF}, the amount of Na-deficiency between the layers of monoclinic \NCTO~is significantly less than that of the hexagonal polymorph -- a direct result of the change of space group. 
Therefore, structural disorders such as bond randomness within the honeycomb layers and stacking faults between them are fewer in the newly synthesized monoclinic polymorph. 

\subsection{\label{subsec:M}Magnetic Characterization}
\begin{figure}
\includegraphics[width=0.46\textwidth]{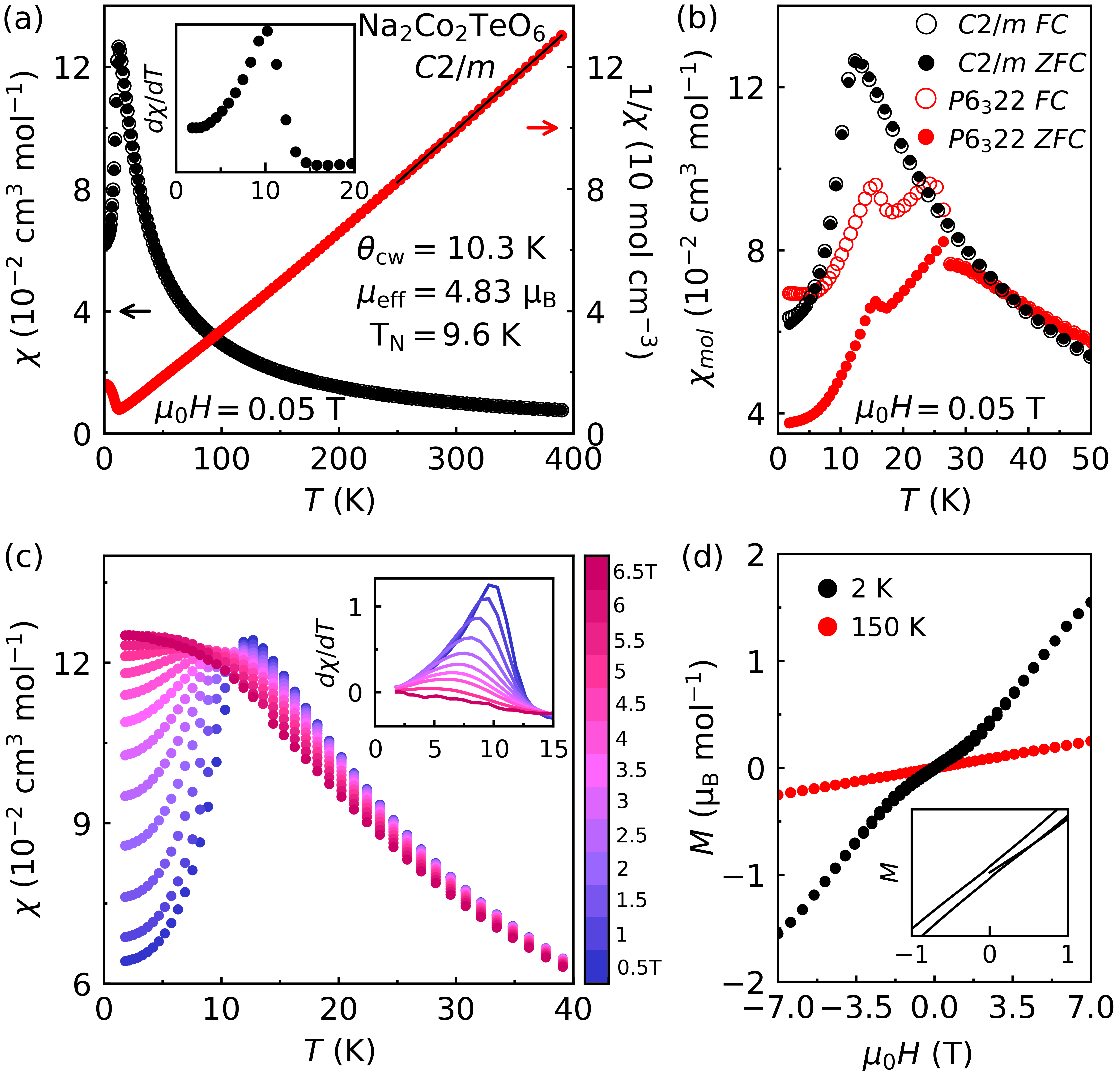}
\caption{\label{fig:MTH}
(a) Magnetic susceptibility per mole Co (black) and inverse susceptibility (red) plotted as a function of temperature. The filled and empty circles correspond to zero-field-cooled (ZFC) and field-cooled (FC) data, respectively. The solid black line is a Curie-Weiss (CW) fit above 250~K.
Inset shows the $d\chi/dT$ curve to identify $T_{\text N}$
(b) Comparison between $\chi(T)$ in the monoclinic and hexagonal polymorphs of \NCTO.
(c) $\chi(T)$ (and $d\chi/dT$ in the inset) at several fields values. 
(d) Magnetization as a function of field at 2 and 150~K. 
Inset shows a weak hysteresis at small fields.
}
\end{figure}
The monoclinic polymorph of \NCTO\ has a single AFM transition characterized by one peak in the susceptibility data $\chi(T)$ without ZFC/FC splitting (Fig.~\ref{fig:MTH}a).
The N\'{e}el temperature $T_{\text{N}}=9.6(6)$~K is determined from the peak in $d\chi/dT$ in the inset of Fig.~\ref{fig:MTH}a.
A comparison between the $\chi(T)$ curves of the monoclinic and hexagonal polymporphs is shown in Fig.~\ref{fig:MTH}b.
The hexagonal polymorph orders at a higher temperature $T_{\text{N}}=27$~K with two spin reorientation transitions at 15 and 5~K (the peak and trough in the ZFC data), which are absent in the monoclinic polymorph~\cite{NaCTO_ACS_2019, nature-NCTO-2021, PhysRevMaterials-NaCTO_2022}.
Figure~\ref{fig:MTH}b also shows the absence (presence) of ZFC/FC splitting in the monoclinic (hexagonal) polymorph indicating the absence (presence) of spin-glass behavior consistent with less (more) Na disorder.
In Appendix~\ref{app:S1S2} we show that a lower quality sample of the monoclinic \NCTO\ has a lower transition temperature (5.9~K instead of 9.6~K) with an upturn at around 3~K, suggesting that the previously reported spin re-orientation transitions in hexagonal \NCTO~\cite{NaCTO_ACS_2019,NCTO_Cm,nature-NCTO-2021,PhysRevMaterials-NaCTO_2022} may be due to an impurity phase of the monoclinic polymorph.

A Curie-Weiss (CW) analysis, $\chi^{-1} = (T - \Theta_{\text{CW}})/ C$, at $T>250$~K in Fig.~\ref{fig:MTH}a yields a CW temperature of $\Theta_{\text{CW}}=+10.3$~K and an effective moment of $\mu_{\text{eff}}= 4.83\,\mu_{\text B}$. 
The positive sign of $\Theta_{\text{CW}}$ in the monoclinic \NCTO\ indicates the presence of FM correlations, unlike in the hexagonal \NCTO\ which has a negative CW temperature ($\Theta_{\text{CW}} = - 8.3$~K). 
In this regard, the behavior of monoclinic \NCTO\ is closer to that of \NCSO\ with $\Theta_{\text{CW}} = +2.2$~K and an AFM order at $T_{\text{N}} = 8.3$~K. 
The similar behavior of these two compounds is likely related to having the same monoclinic structure ($C2/m$).
Table~\ref{tab:CW} summarizes the magnetic parameters of these materials.

\begin{table}
  \caption{\label{tab:CW}Magnetic properties of \NCSO\ and the hexagonal and monoclinic polymorphs of \NCTO.
  }
   \begin{tabular}{l|lll}
   \hline
   \hline
   Material             &  \NCSO                   &  \NCTO                          &  \NCTO                   \\
                        &  Monoclinic              &  Hexagonal                      &  Monoclinic              \\
   \hline
   Space group          &  $C2/m$                  &  $P6_322$                       &  $C2/m$                  \\
   T$_{\text{N}}$       &  $8.3$~K                 &  $27$~K                         &  $9.6$~K                  \\
   $\Theta_{\text{CW}}$ &  $+2.2$~K                &  $-8.3$~K                       &  $+10.3$~K                \\
   $\mu_{\text{eff}}$   &  $5.22$~$\mu_{\text B}$ &  $5.34~\mu_{\text B}$          &  $4.83~\mu_{\text B}$   \\
   S$_{m}$/Co           &  $1.47\,R\ln(2)$         &  $0.70\,R\ln(2)$                &  $0.70\,R\ln(2)$         \\
   Reference            & \cite{NCSO-zigzag-2016}  &  \cite{NCTO_Cm, NaCTO_ACS_2019} &  [this work]             \\
   \hline
   \hline
  \end{tabular}
\end{table}

The effective moment of $4.83\,\mu_{\text B}$ in the monoclinic \NCTO\ is close to the value $4.73\,\mu_{\text B}$ expected from a high-spin $3d^{7}$ system with $S=3/2$ and $L_{\text{eff}}=1$ with unquenched orbital moment ($\text{g}=1.6$ instead of $2$). 
In contrast, the effective moments of hexagonal \NCTO\ ($5.34\,\mu_{\text B}$) and \NCSO\ ($5.22\,\mu_{\text B}$) listed in Table~\ref{tab:CW} are closer to the value $5.92\,\mu_{\text B}$ expected from a spin-only state with quenched orbital moment ($\text{g}=2$)~\cite{NaCTO_ACS_2019}. 
Thus, the effect of SOC seems to be stronger in the title compound compared to its counterparts. 

Figure~\ref{fig:MTH}(c) shows that $T_{\text N}$, defined as the peak in $d\chi/dT$, is suppressed by an external magnetic field of 6~T.
A similar behavior is observed in the hexagonal polymorph, where the suppression of $T_{\text N}$ happens at 9~T~\cite{nature-NCTO-2021}.
Such a behavior is reminiscent of the field-induced quantum paramagnetic phase proposed for $\alpha$-RuCl$_3$~\cite{banerjee_excitations_2018}.  

Figure~\ref{fig:MTH}(d) shows magnetization curves below and above $T_{\text{N}}$ in the monoclinic \NCTO.
At 2~K, $M(\mu_0H=7~\text{T})$ reaches $1.6 \mu_{\text B}$, which is close to the local moment found in neutron diffraction (see below). 
The inset of Fig.~\ref{fig:MTH}(d) shows a weak hysteresis at 2~K for $H<3$~T, evidence of a finite FM component and competing FM/AFM interactions. 
This is consistent with the observed positive $\Theta_{\text{CW}}$ despite AFM ordering (Table~\ref{tab:CW}) as well as the c-type zigzag AFM order found by neutron scattering (see below). 
    
\subsection{\label{subsec:HC}Heat Capacity}
\begin{figure}
\includegraphics[width=0.46\textwidth]{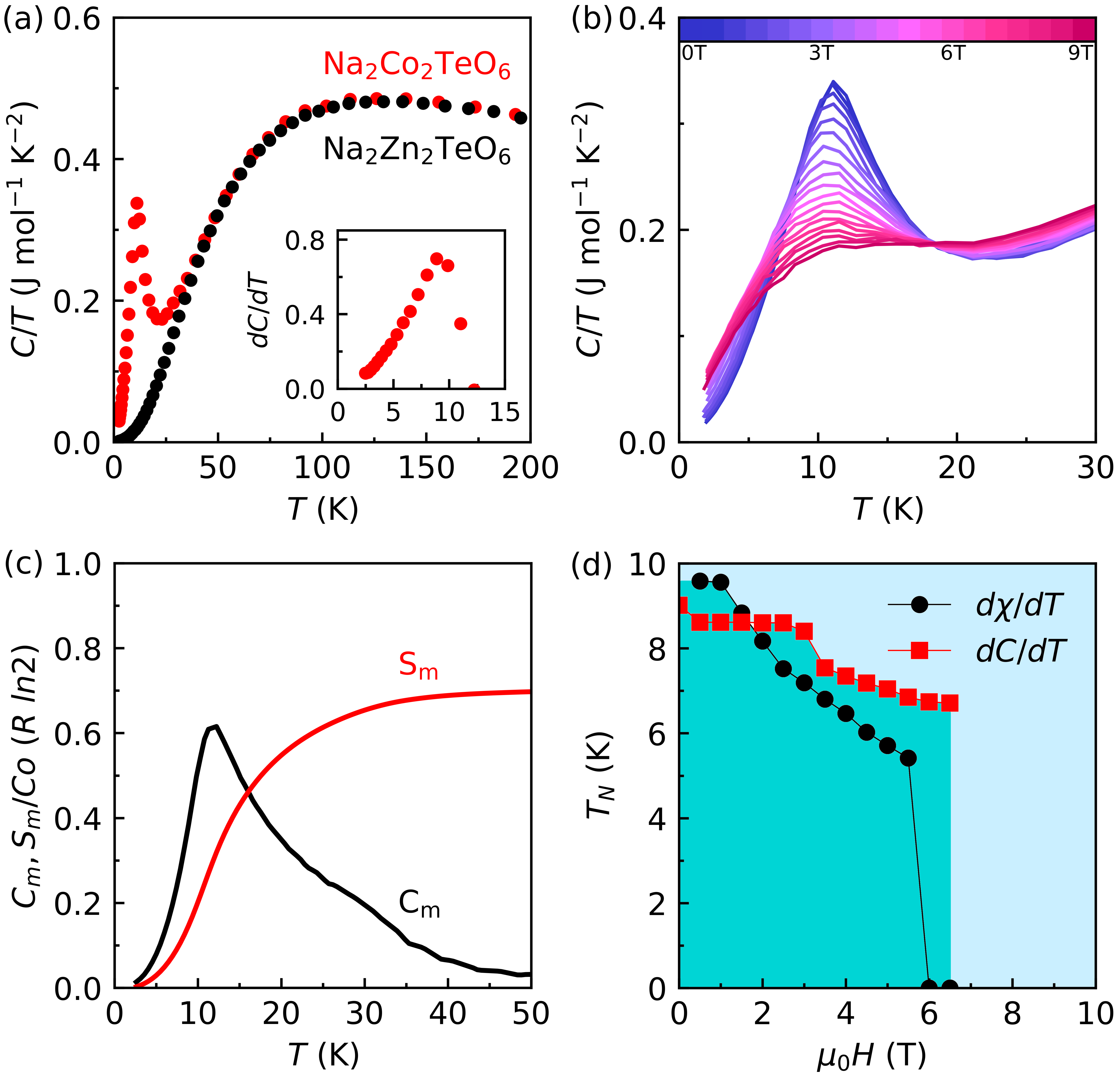}
\caption{\label{fig:CTH}
(a) Heat capacity divided by temperature ($C/T$) per mole Co or Zn plotted as a function of temperature for monoclinic \NCTO~(red) and \NZTO (black). 
The black data is multiplied by 0.95 to correct for the mass difference between Co and Zn. 
Inset shows $dC/dT$ at zero field to determine $T_{\text N}$. 
(b) Magnetic heat capacity ($C_{m}$) and entropy ($S_{m}$) of monoclinic \NCTO~in units of $R\ln(2)$ as a function of temperature.
(c) $C/T$ per mole Co as a function of temperature at different magnetic fields.
(d) Suppression of $T_{\text N}$ with increasing field according to $d\chi/dT$ and $dC/dT$ data. 
}
\end{figure}
Similar to the magnetic susceptibility data, a single peak is observed at 12~K in the heat capacity of monoclinic \NCTO\ due to AFM ordering (Fig.~\ref{fig:CTH}a).
The low-temperature spin reorientation transitions found in the hexagonal \NCTO\ are absent in the monoclonic polymorph according to both magnetic susceptibility and heat capacity data (Figs.~\ref{fig:MTH}a and \ref{fig:CTH}a).
The peak in $dC/dT$ in the inset of Fig.~\ref{fig:CTH}a is used to evaluate $T_{\text N}=9.6(6)$~K consistent with the value reported from $d\chi/dT$ in the inset of Fig.~\ref{fig:MTH}a.
The lower $T_{\text N}$ in the monoclinic polymorph (9.6~K) compared to hexagonal polymorph (27~K) indicates enhanced magnetic frustration due to the change of crystal symmetry (Fig.~\ref{fig:CIF}).

To isolate the magnetic heat capacity, we synthesized monoclinic \NZTO\ (a non-magnetic analog of the title compound) and measured its purely phononic heat capacity (black data in Fig.~\ref{fig:CTH}a). 
After subtracting the phonon background, the magnetic heat capacity ($C_m/T$) is plotted in units of $R\ln(2)$ per mole Co in Fig.~\ref{fig:CTH}b (black curve). 
Also, the magnetic entropy is calculated by numerical integration using $S_{m} = \int(C_{m}/T)dT$ and plotted in Fig.~\ref{fig:CTH}b (red curve).
It reaches $70\%$ of $R\ln(2)$, which is the expected molar entropy per Co$^{2+}$ for the theoretically predicted $\Gamma_{7}$ doublet (pseudo-spin 1/2)~\cite{Motome_2020}. 
Releasing 70$\%$ of this amount across the AFM transition could be due to either an incomplete phonon subtraction or considerable fluctuations of the pseudo-spin $1/2$ degrees of freedom above $T_{\text N}$. 
Table~\ref{tab:CW} compares the magnetic entropy of monoclinic \NCTO\ with its hexagonal polymorph and the isostructural system \NCSO.

Figure~\ref{fig:CTH}c shows that the AFM transition is suppressed gradually by applying a magnetic field.
Using the peaks in both $dC/dT$ and $d\chi/dT$, a temperature-field phase diagram is constructed in Fig.~\ref{fig:CTH}d that shows the suppression of the AFM order at 6~T.
The measured $C/T$ as a function of temperature shows similar behavior to the magnetic susceptibility and displays a suppression of the AFM peak with increasing field. However, in contrast to the complete change of behavior seen in $\chi$ at 6 T, the $C/T$ data still shows a residual peak up to 9 T. Similar suppression of the AFM peak has been reported for \NCTO~\cite{NCTO_Cm}.

\subsection{\label{subsec:NPD}Neutron Powder Diffraction}
\begin{figure}
\includegraphics[width=0.46\textwidth]{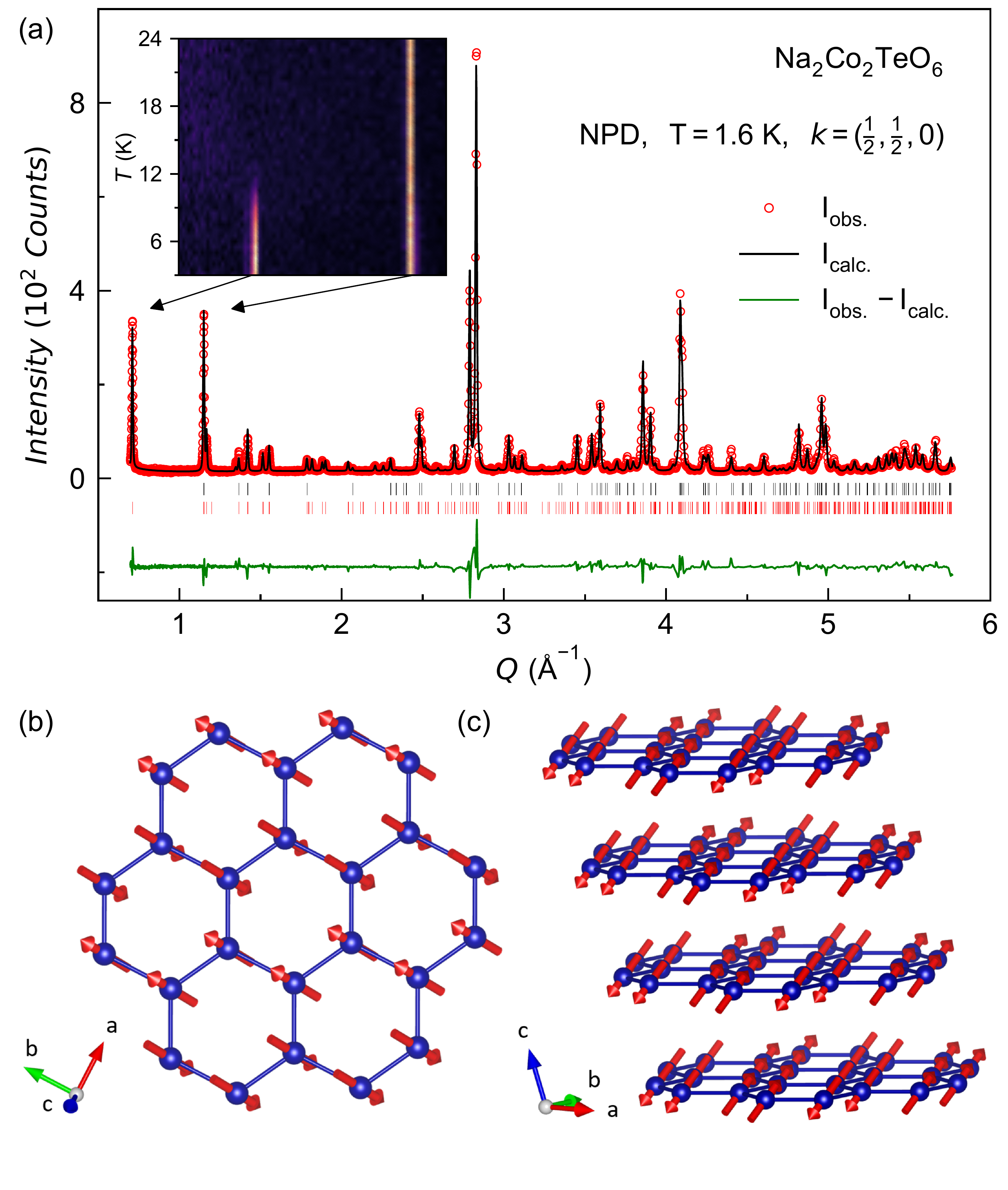}
\caption{\label{fig:NPD}
(a) Neutron powder diffraction (NPD) pattern of the monoclinic \NCTO\ at $T<T_{\text N}$ modeled by a zigzag magnetic structure visualized in the bottom panels. 
(b) The spins are predominantly in the $bc$ plane with $37^\circ$ out-of-plane canting. 
(c) The inter-layer coupling is FM.
}
\end{figure}
To determine the nuclear and magnetic structures, NPD profiles were collected at 100~K (Fig.~\ref{fig:XRD}b) and 1.6~K (Fig.~\ref{fig:NPD}a). 
The black and red ticks in Fig.~\ref{fig:NPD}a mark the positions of the nuclear and magnetic Bragg peaks, the latter of which appears at $T<T_{\text N}$.
The inset of Fig.~\ref{fig:NPD}a compares a temperature independent nuclear Bragg peak at $Q=1.2$~\AA\ to a temperature dependent magnetic Bragg peak at $Q=0.7$~\AA\ with growing intensity below $T_{\text N}$.

The magnetic peaks in Fig.~\ref{fig:NPD}a are indexed by the commensurate propagation vector $\bm{k} = (\frac{1}{2}, \frac{1}{2}, 0)$. 
A magnetic symmetry analysis based on the structural space group $C2/m$ gives two magnetic maximal subgroups corresponding to the zigzag and stripy AFM orders within the honeycomb layers with FM coupling between the layers.
However, the magnetic refinement for the zigzag order produces a higher quality fit than the stripey configuration (Appendix~\ref{app:NPD}).
Thus, the magnetic subgroup that best represents the experimental data is $P_s\overline{1}$ (Irrep: $mV_{1^-}$) which describes a zigzag AFM order within the layers and FM coupling between them (Figs.~\ref{fig:NPD}b,c). 
The non-vanishing $\chi(T)$ when $T\to 0$ and positive $\Theta_{\text{CW}}$ in Fig.~\ref{fig:MTH}a are consistent with such a magnetic structure.

A refinement of the moment size in the zigzag structure gives $\mu = (0.48(15), 1.50(15), 1.18(16))\,\mu_{\text B}$ suggesting that the spins lie primarily along the $b$ axis with $37^\circ$ canting out of plane (Figs.~\ref{fig:NPD}b,c).
The magnetic moment per Co$^{2+}$ from this refinement is $1.83~\mu_{\text B}$ which can be understood by considering the high-spin configuration ($^4F$) of the $3d^7$ orbitals which splits into two triplets and a singlet ($^4F \to 2^4T+^4A$) under the octahedral CEF~\cite{lines_magnetic_1963}.
The lowest energy triplet $^4T$ has an orbital angular momentum $L=-\frac{3}{2}L_{\text{eff}}=-\frac{3}{2}\times 1$ and spin $\frac{3}{2}$ leading to a total moment $\langle m \rangle=2S+L=2(\frac{3}{2})-\frac{3}{2}=\frac{3}{2}$.
This is close to but slightly lower than the observed moment of $1.83~\mu_{\text B}$.
The small difference is likely due to the trigonal distortion which is ignored in the first-order analysis.


\section{\label{sec:CONCL}Conclusion}
The results presented here highlight the interplay between the structural symmetry and magnetic properties in Kitaev magnets. 
Although both polymorphs of \NCTO\ have identical Co-Te honeycomb layers, the magnetic properties of the monoclinic \NCTO~are markedly more similar to the isostructural \NCSO~in space group $C2/m$ than to its hexagonal polymorph in the space group $P6_{3}22$. 
Both monoclinic \NCTO~and \NCSO~have a positive CW temperature and a single AFM transition with evidence of anisotropic interactions and magnetic frustration. 
These results show the importance of lattice symmetry considerations in the ongoing search for an ideal Kitaev candidate material.

\section{Acknowledgments}
The authors thank S.~Nagler and Y.~Ran for fruitful discussions.
The work at Boston College was supported by the National Science Foundation under award number DMR-2203512.
A portion of this research used resources at Spallation Neutron Source, a DOE Office of Science User Facility operated by the Oak Ridge National Laboratory.

\appendix
\section{\label{app:RF}Rietveld Refinement}
\begin{table}
  \caption{\label{tab:T1}Unit cell parameters of \NCTO\ and quality factors of the PXRD Rietveld refinement at room temperature.
  }
  \begin{tabular}{|cc|cc|}
  \hline
  \multicolumn{2}{|c|}{ Unit cell parameters }  &       \multicolumn{2}{c|}{ Refinement parameters } \\
  \hline
   Space Group              & $C2/m$            &  Parameters                     & 20               \\
   $a$  (\AA)               & $5.33225(6)$      & $R_{\mathrm {Bragg}}$ (\%)      & 6.92             \\
   $b$  (\AA)               & $9.20808(8)$      & $R_{\mathrm {F}}$ (\%)          & 5.57             \\
   $c$ (\AA)                & $5.80718(8)$      & $R_{\mathrm {exp}}$ (\%)        & 5.38             \\
   $\beta$  ($^\circ$)      & $108.90837(88)$   & $R_{\mathrm {p}}$ (\%)          & 5.72             \\
   V (\AA$^3$)              & $269.745$         & $R_{\mathrm {wp}}$ (\%)         & 7.69             \\
   $Z$                      & $2$               & $\chi^2$                        & 2.04             \\
   $\rho$~(gr\,cm$^{-3}$)   & $4.770$           & $T$~(K)                         & 295              \\
   \hline
   \end{tabular}
\end{table}
\begin{table}
  \caption{\label{tab:T2}Atomic coordinates, site occupancies, and isotropic Debye-Waller factors from NPD Rietveld refinement of \NCTO\ in space group $C2/m$ at 100~K.
  }
   \begin{tabular}{|ccccccc|}
   \hline
   atom                      &  site  &   $x$     &    $y$    &   $z$    &   occ.       &    $B_{\textrm{iso}}$~(\AA$^2$)   \\
   \hline
   Na1                       &  $4h$  & 1/2       & 0.32818     & 1/2        &  0.700       &    0.014   \\
   Na2                       &  $2d$  & 0         & 1/2         & 1/2        &  0.600       &    0.014   \\
   Co1                       &  $4g$  & 0         & 0.66923     & 0          &  1.000       &    0.007   \\
   Te1                       &  $2a$  & 0         & 0           & 0          &  1.000       &    0.0002  \\
   O1                        &  $8j$  & 0.28060   & 0.34569     & 0.80303    &  1.000       &    0.006   \\
   O2                        &  $4i$  & 0.26474   & 1/2         & 0.19231    &  1.000       &    0.006   \\
   \hline
  \end{tabular}
\end{table}
A co-refinement of PXRD and NPD patterns was used to accurately solve the crystal structure of monoclinic \NCTO.
The unit cell parameters from the PXRD Rietveld refinement are summarized in Table~\ref{tab:T1}. 
Since neutron diffraction is more reliable in determining the oxygen positions, the atomic coordinates, Wyckoff-site occupancies, and Debye-Waller factors are reported from the NPD refinement in Table~\ref{tab:T2}.
Since Na, Co, Te, and O have sufficiently different atomic form factors for neutron diffraction, the chemical composition of \NCTO\ was reliably determined from the NPD refinement.

\section{\label{app:S1S2}Good-quality vs. poor-quality sample}
\begin{figure}
\includegraphics[width=0.46\textwidth]{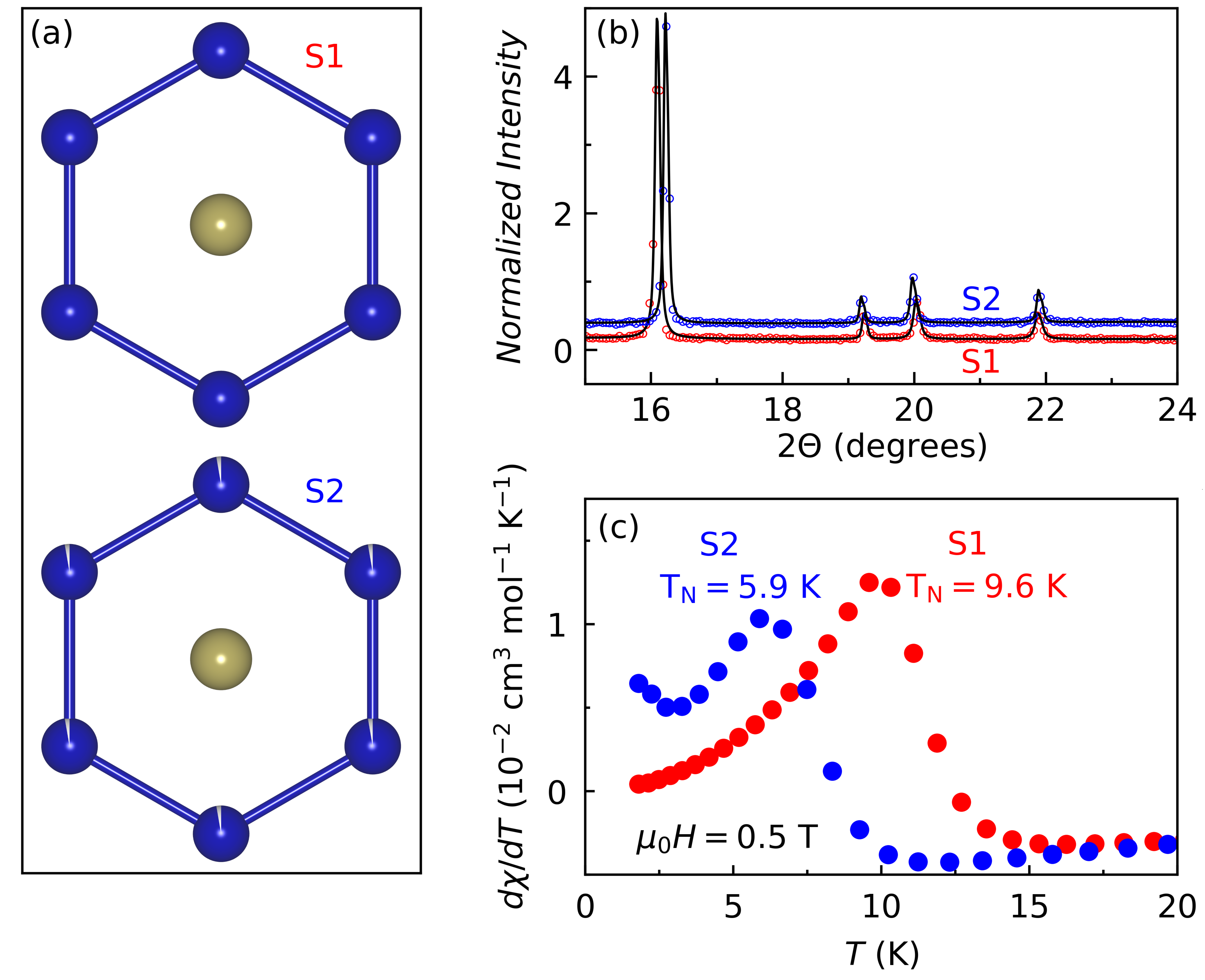}
\caption{\label{fig:S1S2}
(a) A good-quality sample (S1) does not have cobalt deficiency unlike poor-quality (S2) sample.
(b) PXRD pattern of the poor-quality sample (S2) shows a shift of the first peak to the right indicating a larger $c$-axis due to weaker inter-layer bonding.
(c) The magnetic transition is reduced from 9.6 to 5.9~K in the poor-quality sample.
}
\end{figure}
The quality of \NCTO\ samples varies based on the amount of excess Na$_2$CO$_3$ and the temperature and duration of the synthesis.
A common problem in poor-quality samples is cobalt deficiency that is correlated with excess sodium between the layers (to maintain charge neutrality).
Figures~\ref{fig:S1S2}a,b show the results of the PXRD refinements in a good (S1) versus poor (S2) quality sample.
The good quality sample (S1) has less sodium between the layers and no cobalt deficiency.
The poor-quality sample (S2) has more sodium atoms between the layers which strengthen the inter-layer bonds and shorten the $c$-axis.
Thus, the first Bragg peak in Fig.~\ref{fig:S1S2}b is shifted to the right in S2 compared to S1.

Due to cobalt deficiency, $T_{\text N}$ is shifted to a lower temperature in the poor-quality sample (S2) as seen in Fig.~\ref{fig:S1S2}c.
Note that the $T_{\text N}$ reduction in S2 is due to disorder; it is not an evidence of increasing proximity to the Kitaev spin liquid phase.
Also, there is an upturn in $\chi(T)$ of S2 at 3~K similar to the upturn observed in Fig.~\ref{fig:MTH}b in the hexagonal polymorph.
It is likely that this upturn is due to disorder (Co deficiency) in the monoclinic phase and it shows up in hexagonal samples that are contaminated with a small amount of a parasitic monoclinic phase. 

\section{\label{app:NPD}Neutron Diffraction}
\begin{figure}
\includegraphics[width=0.46\textwidth]{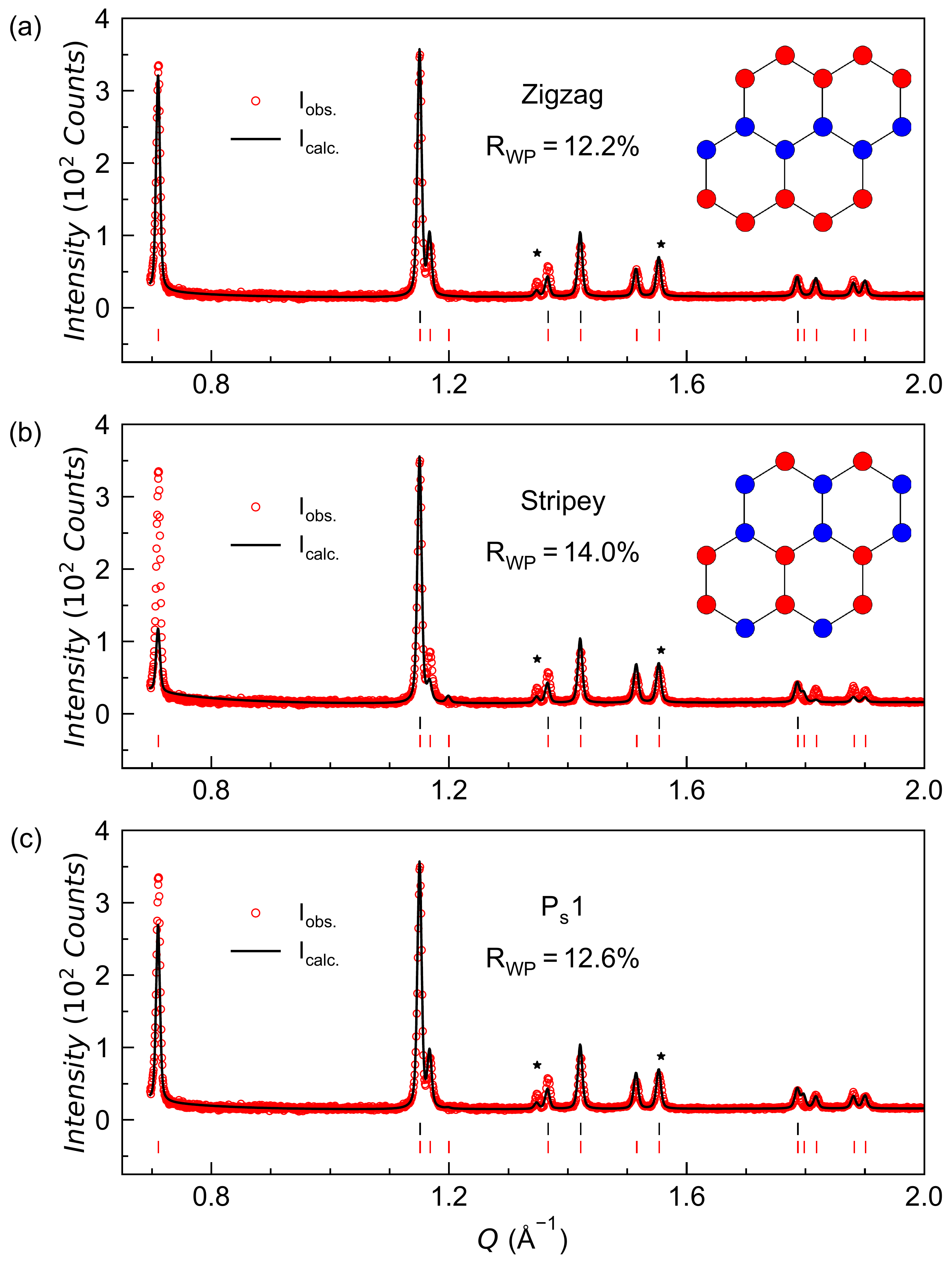}
\caption{\label{fig:NPD2}
Comparison between the magnetic Rietveld refinements of the NPD pattern using three different models.
(a) Maximal magnetic space group $P_s\overline{1}$ with zigzag order, which gives the best fit and amoment of $1.83\,\mu_{\text B}$ per Co$^{2+}$.
(b) Maximal magnetic space group $P_s\overline{1}$ with stripy order, which gives the worst fit.
(c) Magnetic subgroup $P_s1$ with zigzag order but in a lower symmetry magnetic structure. The fit quality is worse than in panel (a).
The red and blue circles in the insets represent anti-parallel spins.
The poor fit quality of the stripy model leads to a larger weighted profile factor $R_{\text WP}$. 
The asterisks mark the positions of Co$_3$O$_4$ impurity peaks.
}
\end{figure}
A symmetry analysis of the $\bm{k}=(\frac{1}{2}, \frac{1}{2}, 0)$ wavevector in the structural space group $C2/m$ of \NCTO\ gives two magnetic models belonging to the maximal magnetic space group $P_s\overline{1}$. 
The irreducible representations of these two magnetic models are $mV_1^-$ and $mV_1^+$ corresponding to the zigzag and stripy orders, respectively.
The Rietveld refinement for both magnetic structures are shown in Fig.~\ref{fig:NPD2}.
Whereas the zigzag model produces a good fit quality, the stripy model does not fit the data properly as seen in Fig.~\ref{fig:NPD2}.
For example, the large Bragg peak at $Q=0.7$ (\AA$^{-1}$) and the small peaks near 1.8 and 1.9 (\AA$^{-1}$) are fitted poorly in the stripy model.
We found a small amount (2\% volume fraction) of Co$_3$O$_4$ impurity in our samples.
The peaks corresponding to this impurity are marked by asterisks in Fig.~\ref{fig:NPD2}. 

It is also possible to refine the NPD pattern in a lower symmetry space group $P_s1$ (irrep. $mV_1^+$) which allows four different Co moment sites, which we constraint to have the same size.
The refinement in this model, which also gives a zigzag in-plane ordering but with $26^\circ$ out-of-plane canting, is presented in Fig.~\ref{fig:NPD2}c.
This model produces a lower quality fit than the first zigzag model in Fig.~\ref{fig:NPD2}a.
It also gives a total moment of $2.91\,\mu_{\text B}$ which is considerably higher than the expected moment from the doublet ground-state ($1.5\,\mu_{\text B}$) and should produce twice the magnetic entropy shown in Fig.~\ref{fig:CTH}b.
Thus, the model that best describes the behavior of the title compound is the zigzag model presented in Figs.~\ref{fig:NPD} and \ref{fig:NPD2}a.


\bibliography{Dufault_12may2023}

\end{document}